%% file: main.tex
\documentclass[%
superscriptaddress,
prl,
aps,
amsmath,amssymb,floatfix,
reprint,%
twocolumn,%
longbibliography,
]{revtex4-1} 

\usepackage{graphicx}
\usepackage{dcolumn}
\usepackage{bm}
\usepackage{amssymb}
\usepackage{multirow}
\usepackage{amsmath}
\usepackage{xcolor}

\newcommand{\1}{\begin{equation}}
\newcommand{\2}{\end{equation}}
\newcommand{\ea}{\begin{eqnarray}} 
\newcommand{\ee}{\end{eqnarray}}

\newcommand{\e}[1]{{ {\rm e}^{#1}  }}

\newcommand{\Intd}{\mathrm{d}}
\newcommand{\vect}[1]{\boldsymbol{#1}}

\usepackage[colorlinks=true,citecolor=cyan, linkcolor=blue]{hyperref}

\begin{document}
\title{Hydrodynamics can Determine the Optimal Route for Microswimmer Navigation}

\date{\today}

\pacs{..}

\author{Abdallah Daddi-Moussa-Ider}
 \affiliation{Institut f\"{u}r Theoretische Physik II: Weiche Materie, Heinrich-Heine-Universit\"{a}t D\"{u}sseldorf, D-40225 D\"{u}sseldorf, Germany}
 \author{Hartmut L\"owen}
 \affiliation{Institut f\"{u}r Theoretische Physik II: Weiche Materie, Heinrich-Heine-Universit\"{a}t D\"{u}sseldorf, D-40225 D\"{u}sseldorf, Germany
 \\e-mail: liebchen@fkp.tu-darmstadt.de}
 \author{Benno Liebchen}
 \affiliation{Institute of Condensed Matter Physics, Technische Universit\"{a}t Darmstadt, D-64289 Darmstadt, Germany}

\begin{abstract}
\textbf{Abstract:}
\\Contrasting 
the well explored problem on how to steer a macroscopic agent like an airplane or a moon lander to optimally reach a target, 
the optimal navigation strategy for microswimmers experiencing hydrodynamic interactions with walls and obstacles is still unknown.
Here, we systematically explore this problem and show that the characteristic microswimmer-flow-field  
crucially influences the required navigation strategy to reach a target fastest. The resulting optimal trajectories 
can have remarkable and non-intuitive shapes, which 
qualitatively differ from those of dry active particles or motile macroagents.
Our results provide generic insights into the role of hydrodynamics and fluctuations on optimal navigation at the microscale and
suggest that microorganisms might have survival advantages when strategically controlling their distance to remote walls. 
\end{abstract}


\maketitle

\section*{Introduction}
The quest on how to navigate or steer to optimally reach a target is important e.g.\ for airplanes to save fuel while facing complex wind patterns on their way to a remote destination, 
or for the coordination of the motion of the parts of 
a space-agent to safely land on the moon.
These classical problems are well-explored and are usually solved using optimal control theory \cite{kirk2004}. 
Likewise, navigation and search strategies are frequently encountered in a plethora of biological systems, including the foraging of animals for food,~\cite{viswanathan11}
or of T cells searching for targets to mount an immune response~\cite{fricke16}.
Very recently there is a growing interest also in optimal navigation problems and search strategies~\cite{muinos2018,Yang2018,Yang2019,liebchen2019,sturmer2019,biferale2019} of microswimmers~\cite{lauga09,elgeti2015,lauga16, lauga2020fluid} and ``dry'' active Brownian 
particles~\cite{romanczuk2012,bechinger2016,zottl16,cates2015, gompper2020}. 
These active agents can self-propel in a 
low-Reynolds-number solvent, and might play a key role in tomorrow's nanomedicine as recently popularized e.g.\ in~\cite{harari2016}. 
In particular, they might become useful for the 
targeted delivery of genes~\cite{qiu2015} or drugs~\cite{park2017, wang12} and other cargo~\cite{ma2015,demirors2018} to a certain target (e.g.\ a cancer cell) through our blood vessels, requiring them to 
find a good, or ideally optimal, path towards the target avoiding e.g.\ obstacles and unfortunate flow field regions. 
In the following, we refer to the general problem regarding the optimal trajectory of a microswimmer which can freely steer but cannot control its speed towards a predefined target (point-to-point navigation) as
``the optimal microswimmer navigation problem''. 

The characteristic differences between the optimal microswimmer navigation problem and conventional optimal control problems for macroagents 
like airplanes, cruise-ships or moon-landers, root in the presence of a low-Reynolds-number solvent in the former problem only. 
They comprise (i) overdamped dynamics (ii) thermal fluctuations and (iii) long-ranged fluid-mediated hydrodynamic interactions with interfaces, walls and obstacles, 
all of which are characteristic for microswimmers \cite{bechinger2016}. 
Notice in particular that the non-conservative hydrodynamic forces which microswimmers experience call for a distinct navigation strategy than the conservative gravitational forces acting e.g.\ on space vehicles.
Recent work has explored optimal navigation problems of dry active particles (and particles in external flow fields) accounting for (i) and partly also for (ii): 
Specifically, 
the very recent 
works \cite{colabrese2017, gustavsson2017,colabrese2018,muinos2018, Yang2018,yoo16,reddy16,reddy18, sturmer2019,biferale2019,alageshan2019,tsang2020self,tsang2020roads} have pioneered the usage of 
reinforcement learning \cite{sutton1998introduction,cichos20,garnier19} e.g. to determine optimal steering strategies of 
active particles to optimally navigate towards a target position~\cite{muinos2018, Yang2018, sturmer2019,biferale2019} or to exploit external flow fields to avoid getting trapped in certain flow structures 
by learning smart gravitaxis~\cite{colabrese2017}. 
Meanwhile, refs.~\cite{Yang2018,Yang2019,yang2020} have used (deep) reinforcement learning to explore microswimmer navigation problems in mazes and obstacle arrays assuming global \cite{Yang2018} or only local \cite{Yang2019} knowledge 
of the environment.
Very recent analytical approaches \cite{liebchen2019,sturmer2019} to optimal active particle navigation 
complement these works and allow testing machine-learned results \cite{biferale2019,sturmer2019}.
(In addition, note that a significant knowledge exists 
on the complementary problem of optimizing body-shape deformation of deformable swimmers with optimal control theory; see e.g. \cite{giraldi2015optimal, alouges2008optimal, alouges2019energy}.)
Despite this remarkable progress in recent years, (iii), and its interplay with (ii), remains an important open problem to understand the optimal microswimmer navigation strategy.   

To fill this gap, in the present work, we systematically explore the optimal microswimmer navigation problem in the presence of walls or obstacles, where hydrodynamic microswimmer-wall interactions are well known to 
occur~\cite{elgeti2010hydrodynamics, li2014hydrodynamic, elgeti2013wall, mozaffari2016self, ibrahim16, mozaffari2018self,shen2018hydrodynamic,elgeti2016microswimmers,laumann2019emerging}, but whose impact on 
optimal microswimmer navigation is essentially unknown. 
Combining an analytical approach with numerical simulations, we find that in the presence of remote obstacles or walls, the shortest path is not fastest for microswimmers, even in the complete absence of external force or flow fields. Thus, unlike dry active particles (or light rays following Fermat's princple), in the presence of remote obstacles microswimmers generically have to take excursions to reach their target fastest. 
In the presence of fluctuations, the ``optimal'' navigation strategy effectively protects microswimmers from fluctutions and can drastically decrease the traveling time as compared to 
where the microswimmer head straight towards the target. 
This offers a novel perspective on the motion of microorganisms near surfaces or interfaces: it 
suggests that microorganisms might have a survival advantage when actively
regulating their distance to remote walls in order to approach a food source via a strategic detour, 
rather than directly heading towards it. 
Besides their possible biological implications, 
our findings might provide a benchmark for future research on optimal navigation strategies of active particles.

\section*{Results and Discussion}
\begin{figure}
	\centering
	\includegraphics[width=0.49\textwidth]{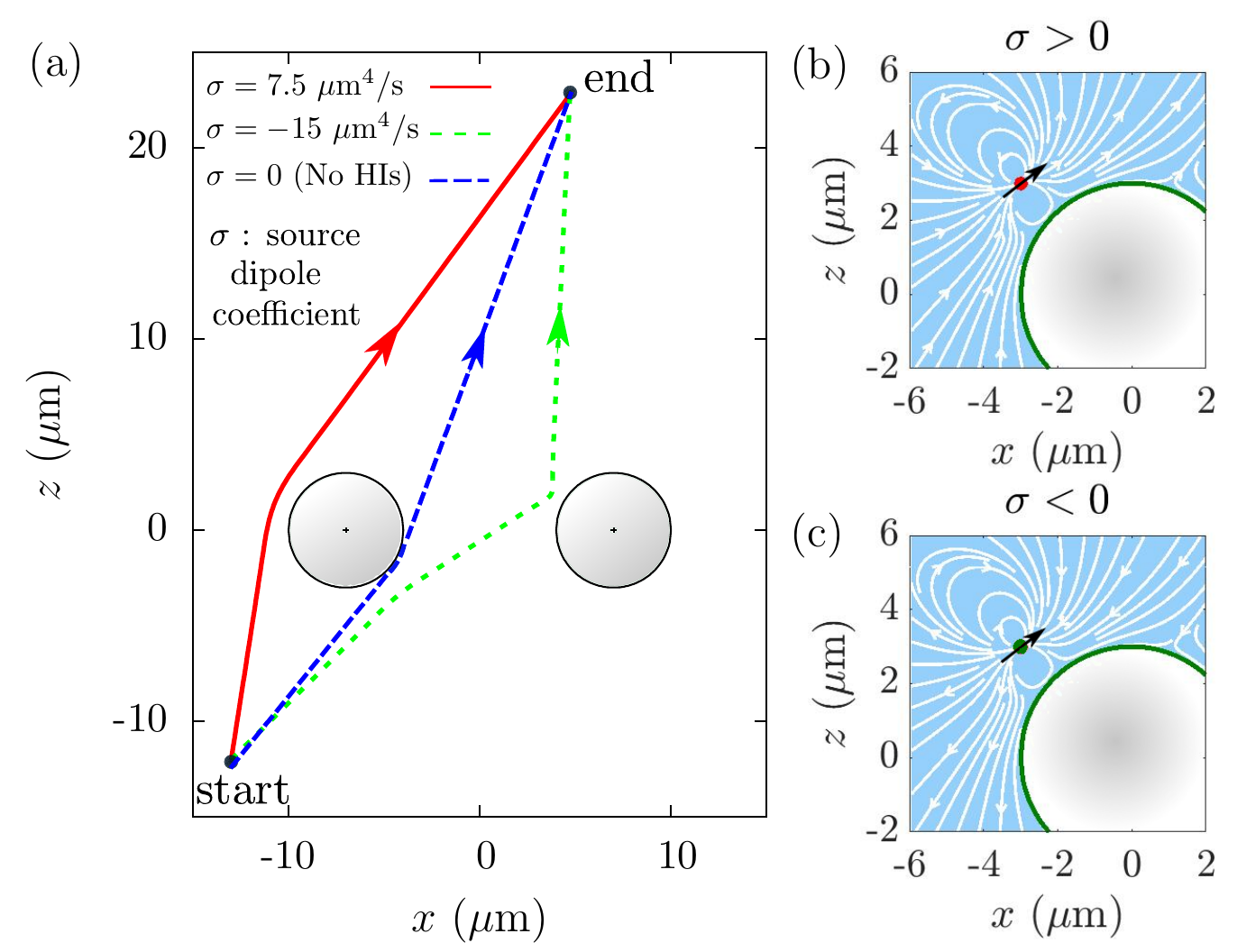}
	\caption{(a) Optimal microswimmer navigation versus conventional optimal navigation in the presence of two spherical obstacles (shown in gray): Curves represent optimal trajectories for microswimmers (red and green) and for a dry active particle (blue), {i.e., in the absence of hydrodynamic interactions (HIs) with obstacles.}
	{The swimmers are micron sized and the swimming trajectories are assumed to take place in the plane passing through the centers of the spherical obstacles.}
	{Here, $\sigma$ is the  source dipole {coefficient} and $x,z$ are spatial coordinates.}
	Panels (b),(c) show the 
	flow field streamlines induced by a source dipole at position $\vect{r}=(-3,3)~\mu$m
	in the presence of a spherical obstacle with radius $R=3~\mu$m.
	{Black arrows indicate the orientation of the swimmers.
	The fluid-mediated hydrodynamic interactions with the spherical boundary induce a deceleration of the swimming agent for $\sigma>0$ leading to larger speeds as the swimmer gets away from the obstacle.
	In contrast to that, hydrodynamic interactions cause an acceleration for $\sigma<0$ yielding an increased speed near the obstacle.}
	}
	\label{fig1}
\end{figure}

Before introducing our detailed model, let us illustrate the consequences of the finding that 
the shortest path is not fastest for microswimmers:
Consider a microswimmer which can freely control its swimming direction (but not its speed) and aims to reach a predefined target in the presence of two 
obstacles (Fig.~\ref{fig1}): While in the absence of hydrodynamics (dry active particle), the shortest path is optimal (blue), an actual microswimmer takes a qualitatively different path to reach its target fastest 
(red and green curves) because it produces a flow field which is reflected by the obstacles and changes its speed. In particular, ``source dipole''
microswimmers (specified below) which 
produce a flow field which slows them down near walls
(Fig.~\ref{fig1}b), take an excursion (red curve in panel a) to avoid the obstacles. Other source-dipole microswimmers which produce an analogous but sign-reversed flow field (Fig.~\ref{fig1}c),
which speeds them up near walls, take a qualitatively different excursion to benefit from the proximity of both obstacles (green curve in panel a). 
More generally, we find that the role of hydrodynamics on optimal microswimmer routes can be subtle and lead to counterintuitive trajectory-shapes: 
while for source-dipole swimmers the sign of the coefficient plays a decisive role for the resulting trajectory-shapes, as have just seen (Fig.~\ref{fig1}), for 
for force-dipole swimmers, we will see that the sign of the coefficient is unimportant for optimal navigation and only the strength of the hydrodynamic interactions matters.

\subsection*{Model}
{Let us consider a self-propelling active particle interacting with a 3D fluctuating environment.
The particle's center of mass evolves as 
$\dot {\vect{r}}(t) = \left(v_x({\vect{r}},\psi, t), v_y({\vect{r}},\psi, t), v_z({\vect{r}},\psi, t)\right) + \sqrt{2D} \, \boldsymbol{\eta}(t)$, wherein $v_x$, $v_y$, and $v_z$ are the components of the deterministic swimming velocity in Cartesian coordinates, which depend on the hydrodynamic swimmer-wall interactions} 
{as well as on the propulsion direction of the swimmer $(\cos \psi\cos\phi, \cos\psi \sin\phi, \sin\psi)$.}
Here, $D$ is the diffusion coefficient which determines the strength of thermal fluctuations; for now, we choose $D=0$ {and consider a two-dimensional motion in the $xz$-plane and thus set $\phi=0$.
We will discuss the effect of fluctuations later.}
Given a predefined initial ${\vect{r}}(t=0)={\vect{r}}_A$ and terminal point ${\vect{r}}(t=T)={\vect{r}}_B$ in the $xz$-plane, we ask for the optimal connecting trajectory minimizing the travel time~$T$, when the swimmer is allowed to steer freely. (This may represent e.g.\ biological swimmers which steer through body shape deformations or synthetic swimmers controlled by external feedback). 
{That is, $\psi(t)$ can be freely chosen so as to minimize the ravel time of the swimmer.}
This is a well-defined optimal control problem determining the optimal trajectory,  
the navigation protocol $\{\psi(t)|t\in [0,T]\}$ and $T$. It resembles classical navigation problems e.g.\ of an airplane, 
which can steer freely and move at a speed which is determined by the wind (assuming some favorable constant engine power). 
Interestingly, however, while for such macroagents or dry active particles
in constant external fields the shortest path is optimal~\cite{born1970principles,liebchen2019}, for microswimmers excursions can pay off, as we will see in the following. 
Note that the considered model can be tested with programmed active colloids \cite{bregulla2014stochastic,lavergne2019group,sprenger2020active} but should also be relevant for biological microswimmers which often swim at an almost constant speed and are able to control their self-propulsion direction on demand. These swimmers are often in contact with (remote) walls or interfaces which drastically influence their overall swimming speed and direction of motion.

{As a model microswimmer, in the following, we employ the multipole description of swimming microorganisms. 
Accordingly, the self-generated flow field is decomposed in the far-field limit, to a good approximation, into a superposition of higher-order singularities of the Stokes equations. 
This model has favorably been verified experimentally for swimming E.\ coli bacteria~\cite{berke08} and Chlamydomonas algae~\cite{drescher2010direct}.
Even though the theory is based on a far-field description of the hydrodynamic flow, it has been demonstrated using boundary integral simulations that, in some cases, the far-field approximation is surprisingly accurate, all the way down to a microswimmer-wall distance of one tenth of a swimmer length~\cite{spagnolie12}.}

\vskip 0.5cm
\textbf{Source dipole swimmers:}
To develop an elementary understanding of optimal microswimmer navigation, let us first consider a source dipole microswimmer (e.g.\ Paramecium or active colloids with uniform surface mobility) aiming to reach 
a target in the presence of a distant hard wall infinitely {extended in the $xy$-plane, for an initial and target position in the $xz$-plane,} 
yielding \cite{spagnolie12} {$v_x = \left[ v_0 - \sigma/\left(4 z^3 \right) \right] \cos\psi$, $v_y = 0$, and $v_z = \left( v_0 - \sigma/z^3 \right) \sin\psi$
wherein} the deviation from $v_0$ is due to hydrodynamic swimmer-wall interactions.

To reduce the parameter space to its essential dimensions we choose the
length unit as $\ell = (\left| \sigma \right| /v_0)^{1/3}$, which represents the swimmer-wall distance at which the swimmer displacement per time due to hydrodynamic interactions and due to self-propulsion become comparable.
For the time unit, we chose the associated time scale $\tau= \left| \sigma\right|^{1/3} /v_0^{4/3}$.
In reduced units, {the noise-free} equations of motion for a source dipole swimmer then read, 
\begin{subequations}\label{sdeq}
	\begin{eqnarray}
	\dot x^\ast &=& \left(1 - \frac{s}{4 {z^\ast}^3}\right) \cos\psi \, , \\
	\dot z^\ast &=& \left(1 - \frac{s}{{z^\ast}^3}\right) \sin\psi \, , 
	\end{eqnarray}
\end{subequations}
where $x^\ast=x/\ell,z^\ast=z/\ell$, and $s=\text{sgn}(\sigma)$.
Accordingly, {by expressing the equations of motion together with the underlying boundary conditions in reduced units,} the optimal microswimmer trajectories only depend on the sign of the source dipole parameter, but not on its strength or on the swimmer speed. 
For microswimmers achieving self propulsion through surface activity (ciliated microorganisms like Paramecium, active colloidal particles with uniform surface mobility) one expects~$\sigma>0$, i.e.\ $s=1$ 
whereas $\sigma<0$ ($s=-1$) applies to some non-ciliated microswimmers with flagella~\cite{mathijssen2015hydrodynamics}.
\vskip 0.5cm 
To solve the optimal navigation problem, 
we first eliminate~$\psi$ from the equations of motion {(Methods)}.
We then {obtain} the travel time as $T^\ast  = \int_{x_A^\ast}^{x_B^\ast} |\dot x^\ast|^{-1} \, {\rm d}x^\ast =  \int_{x_A^\ast}^{x_B^\ast} \mathcal{L}_\mathrm{SD} \left( x^\ast, z^\ast(x^\ast),{z^\ast}^\prime (x^\ast) \right) \, \Intd x^\ast$, wherein $z^\ast(x^\ast)$ represents the path which we optimize {and $\dot{z}^\ast= {z^\ast}' \dot{x}^\ast$ with ${z^*}' = \partial z^\ast/\partial x^\ast$.}
To find the optimal path which minimizes $T^\ast$, we now determine the Lagrangian 
as (see Methods for details) \begin{equation}
\mathcal{L}_\mathrm{SD}^\ast := \left| \dot{x}^\ast \right|^{-1} = 
\left( \frac{1}{\left(1 - \frac{s}{4 {z^\ast}^3}\right)^2} + \frac{{{z^*}^\prime}}{ \left(1 - \frac{s}{{z^\ast}^3}\right)^2} \right)^{1/2}
\end{equation}
and then numerically solve the 
Euler-Lagrange {equation} for $\mathcal{L}_\mathrm{SD}^\ast$ as a boundary value problem, with the boundary conditions $z^\ast(x^\ast_A)=z^\ast_A$ and $z^\ast(x^\ast_B)=z^\ast_B$, using shooting methods. 


\textbf{Force dipole and force quadrupole swimmers:}
Similarly, the translational swimming velocities due to force dipolar hydrodynamic interactions (E.\ coli, Chlamydomonas) with a planar hard wall reads~\cite{berke08, spagnolie12}
$\dot{x} = v_0 \cos\psi + 3\alpha\sin(2\psi)/(8z^2) $  
and $\dot{z} = v_0 \sin\psi + 3\alpha\left[ 1-3\cos(2\psi) \right]/(16z^2) $ 
where~$\alpha$ is the force dipole coefficient. 
In units of $\ell= (|\alpha|/v_0)^{1/2}$ and $\tau=\left|\alpha\right|^{1/2}/v_0^{3/2}$
the equations of motion 
read 
\begin{subequations}\label{fdeq}
	\begin{eqnarray} 
	{\dot x}^\ast &=& \cos \psi + \frac{3 s \sin (2\psi)}{8 {z^\ast}^2} \, ,  \label{fdeqa} \\ {\dot z}^\ast &=& \sin \psi 
	+ {\frac{3 s \left[ 1-3 \cos (2\psi) \right]}{16 {z^\ast}^2}} \, . \label{fdeqb}
	\end{eqnarray}
\end{subequations}
Here, $s=\text{sgn}(\alpha)$ {is the sign} of the singularity coefficient.
After some algebra, as detailed in the Methods section, the resulting Lagrangian follows as
\begin{equation}
	\mathcal{L}_\mathrm{FD} = \left|  \frac{72
	{z^\ast}^2 {z^\ast}^\prime}{r_\pm^2 - 16 {z^\ast}^4 - 27
	} \right| \, , \label{FDLagr}
\end{equation}
where $r_\pm$ are the roots of a lengthy quartic polynomial, the coefficients of which are explicitly known functions of $z^\ast$ and~${z^\ast}^\prime$ (see Methods). 
The optimal swimming trajectories then result again from solving the Euler-Lagrange {equation} as a boundary value problem using shooting methods.
Interestingly, $\mathcal{L}_\mathrm{FD}$ is sign invariant in the force dipole coefficient~$\alpha$, which means that 
pushers and pullers show identical optimal trajectories, albeit they require different navigation strategies for this.
We will discuss this further in the {Results} section. 
\\Finally, we also calculate the Lagrangian for force quadrupole microswimmers as detailed in the {Methods} section. 

\subsection*{Optimal microswimmer trajectories}
\begin{figure}
	\centering
	\includegraphics[width=0.49\textwidth]{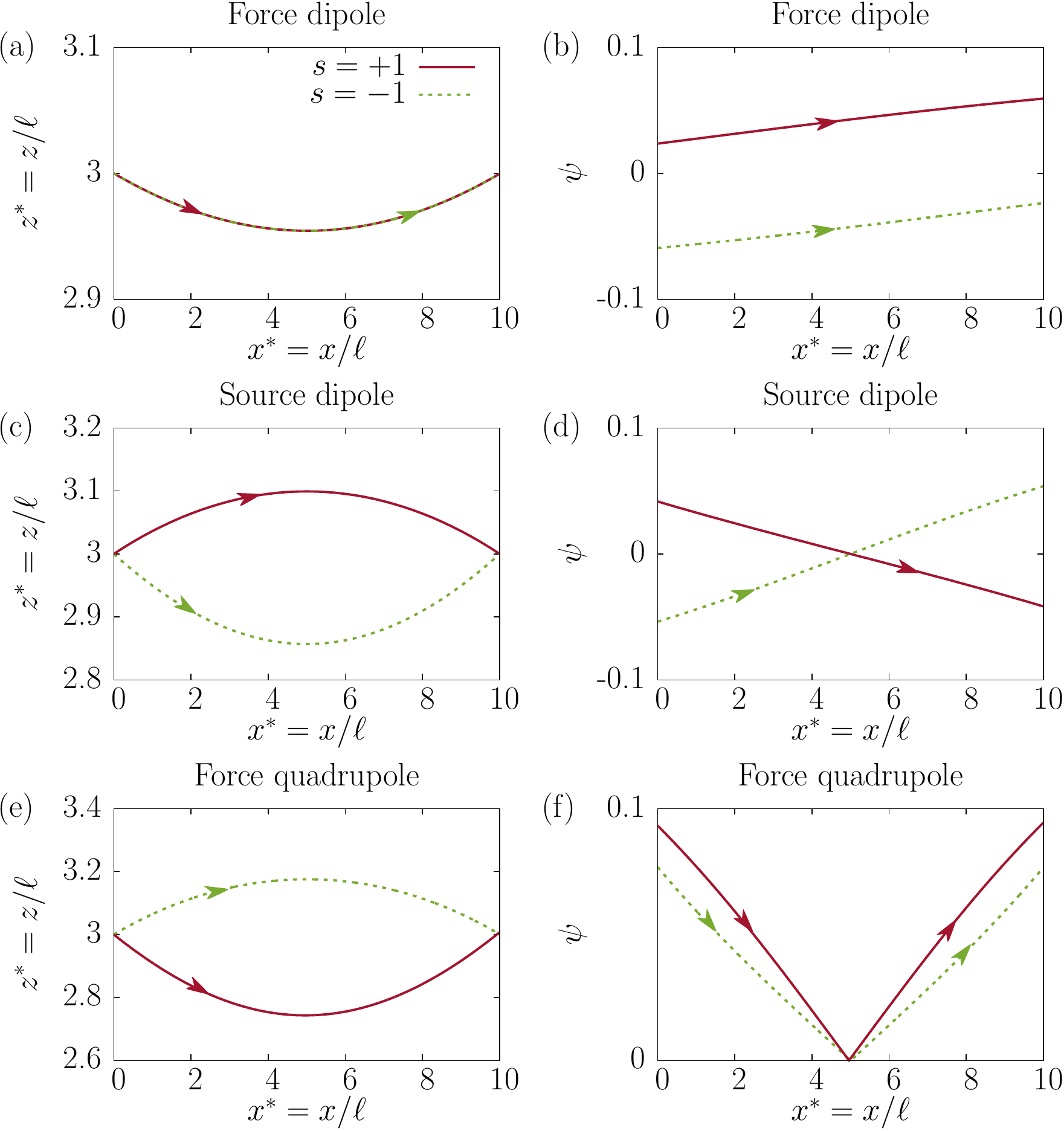}
	\caption{Optimal {2D} microswimmer trajectories minimizing traveling times between {$\vect{r}^\ast_A=(0,0,3)$ and $\vect{r}^\ast_B=(10,0,3)$} {and the corresponding steering angles for force dipole swimmers [(a) and (b)], source dipole swimmers [(c) and (d)], and force quadrupole swimmers [(e) and (f)],} in the presence of a hard and infinitely extended wall.
	{The swimming trajectories take place in the $xz$-plane.}
	{Red solid and green dashed} lines correspond to positive $(s=1)$ and negative $(s=-1)$ singularity coefficient, respectively.
	}
	\label{fig2}
\end{figure}

\textbf{Flat walls:}
As shown in {Fig.~\ref{fig2},} in the presence of an infinitely extended and distant flat wall, we find that all considered swimmers (source and force dipole 
swimmers as well as force-quadrupole swimmers) follow significant excursions to reach the target fastest. That is, the shortest path is not fastest. 
{For instance, panel (c)} of Fig.~\ref{fig2} shows that source dipoles 
with {$s=1$} (which slow down when approaching the wall) follow a parabola bended away from the wall, whereas 
those with {$s=-1$} prefer reducing their distance to the wall which speeds them up. 
{The corresponding steering angles required for the optimal navigation strategy are shown in panel (d).}
In contrast to source dipole swimmers, 
perhaps surprisingly, for force dipole swimmers the shape of the resulting parabola depends only on the force dipole strength but not 
on the sign of the flow field {[panels (a) and (b)].}
This pusher-puller-identity is generic not only for planar walls but also applies for spherical obstacles and obstacle 
landscapes, as can be 
directly seen from the independence of the Lagrangian of $s$ [Eq.~\eqref{FDLagr}]. 
Interestingly, however, the required steering protocol, i.e.\ the temporal evolution of the optimal value of the control variable $\psi(t)$, which the swimmer has to choose to realize the optimal path is different 
for pushers and for pullers {(Fig.~\ref{fig2}b)}. 
Force quadrupolar microswimmers describing small microswimmers with elongated flagella~\cite{mathijssen2015hydrodynamics, daddi18}), can also be solved using the Lagrangian approach; 
{their swimming trajectories and steering angles are presented in panels (e) and~(f), respectively.
The resulting parabolic curves} are bent towards or away from the wall depending on the sign of the force quadrupole coefficient.
  
\begin{figure}
	\centering
	\includegraphics[width=0.4\textwidth]{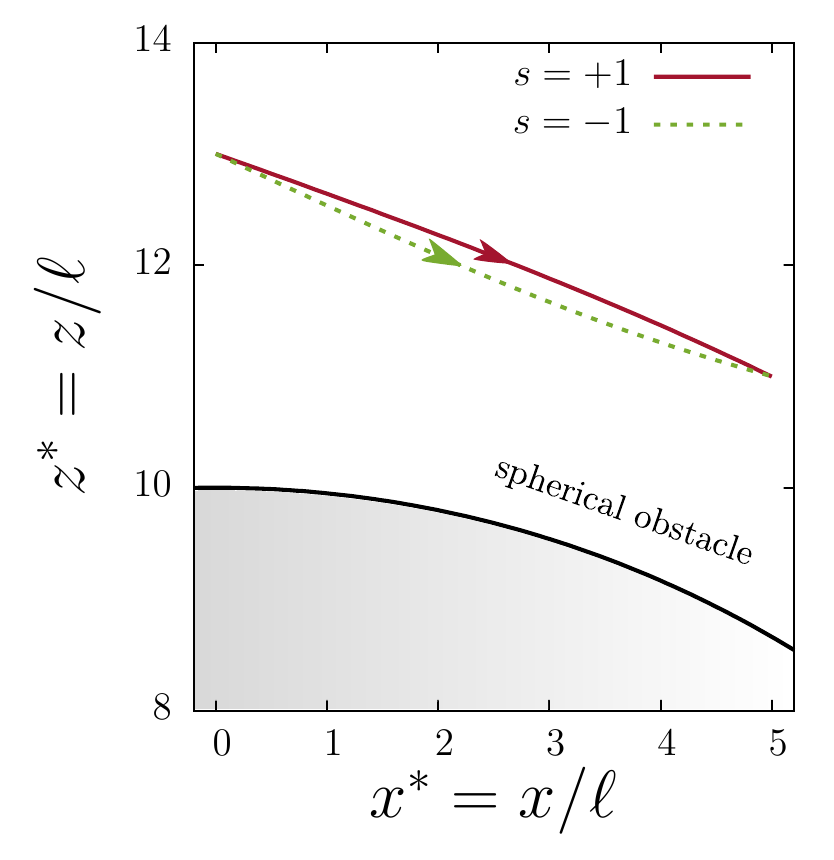}
	\caption{{An exemplary optimal swimming trajectory directed {from (0,0,13) to (5,0,11)} for source dipole microswimmers near a spherical boundary of scaled radius $R^\ast = 10$ positioned at the center of the system of coordinates.}}
	\label{fig3}
\end{figure}

\textbf{Spherical obstacles and complex landscapes:} 
Based on these results we can now understand why hydrodynamic interactions with obstacles can have a drastic impact on the required navigation strategy to cross an obstacle field fastest. 
As shown in Fig.~\ref{fig1} without hydrodynamic interactions the agent 
takes the shortest path (blue curve), whereas a source dipole microswimmer takes qualitatively different path, which {depends} on the sign of the singularity coefficient. 
This is because source dipole swimmers with $s=1$ ($\sigma>0$) tend to avoid flat walls ({red} solid lines in {Fig.~\ref{fig2}c}) as well as spherical obstacles ({red} solid lines in {Fig.~\ref{fig3}}) and are faster when staying at a certain distance to the obstacles. This explains that the red path, which is longer than the blue one, is faster than the blue one for $s=1$-source dipole swimmers in Fig.~\ref{fig1}. Conversely, swimmers  
with $s=-1$ ($\sigma<0$) speed up near flat or spherical walls ({green} dashed lines in {Figs.~\ref{fig2}c and~\ref{fig3}}), which explains why they manage to cross the obstacle field in Fig.~\ref{fig1} faster when following the green trajectory than the shorter blue trajectory. 
These observations demonstrate that the optimal microswimmer navigation strategy qualitatively differs from the optimal strategy of dry active particles or macroagents.

\vskip 0.5cm
\textbf{Fluctuating environments:}
In the world of microswimmers, fluctuations often play an important role. 
Besides Brownian noise which significantly displaces small biological microorganisms or active colloids on their way to a target, 
steering errors (or delay effects \cite{khadem2019delayed}) can effectively lead to fluctuations even in larger microswimmers. 
We now exemplarically consider source dipole microswimmers and set $D^\ast := D/\left(\ell^2/\tau\right) = D/ \left(v_0^{2/3} |\sigma|^{1/3}\right) \neq 0$ assuming that $D^\ast$ does not depend on space for simplicity. (Note that 
accounting for rotational diffusion e.g.\ to represent imperfect steering, does not qualitatively change the following 
results.) 
Let us now compare the following two different navigation strategies: 
The first one, which we call the ``straight swimming strategy''
is to steer straight towards the target at each instant of time. 
An alternative strategy is 
to re-calculate the optimal path of the underlying noise-free problem at each point in time,
using the present position as a starting point, and to steer in the correspondingly determined direction. 
We refer to this as the ``optimal swimming strategy''. 
While the latter strategy is of course expected to be better at weak noise,
for strong noise, 
one might expect the opposite. 

\begin{figure}
\includegraphics[width=0.48\textwidth]{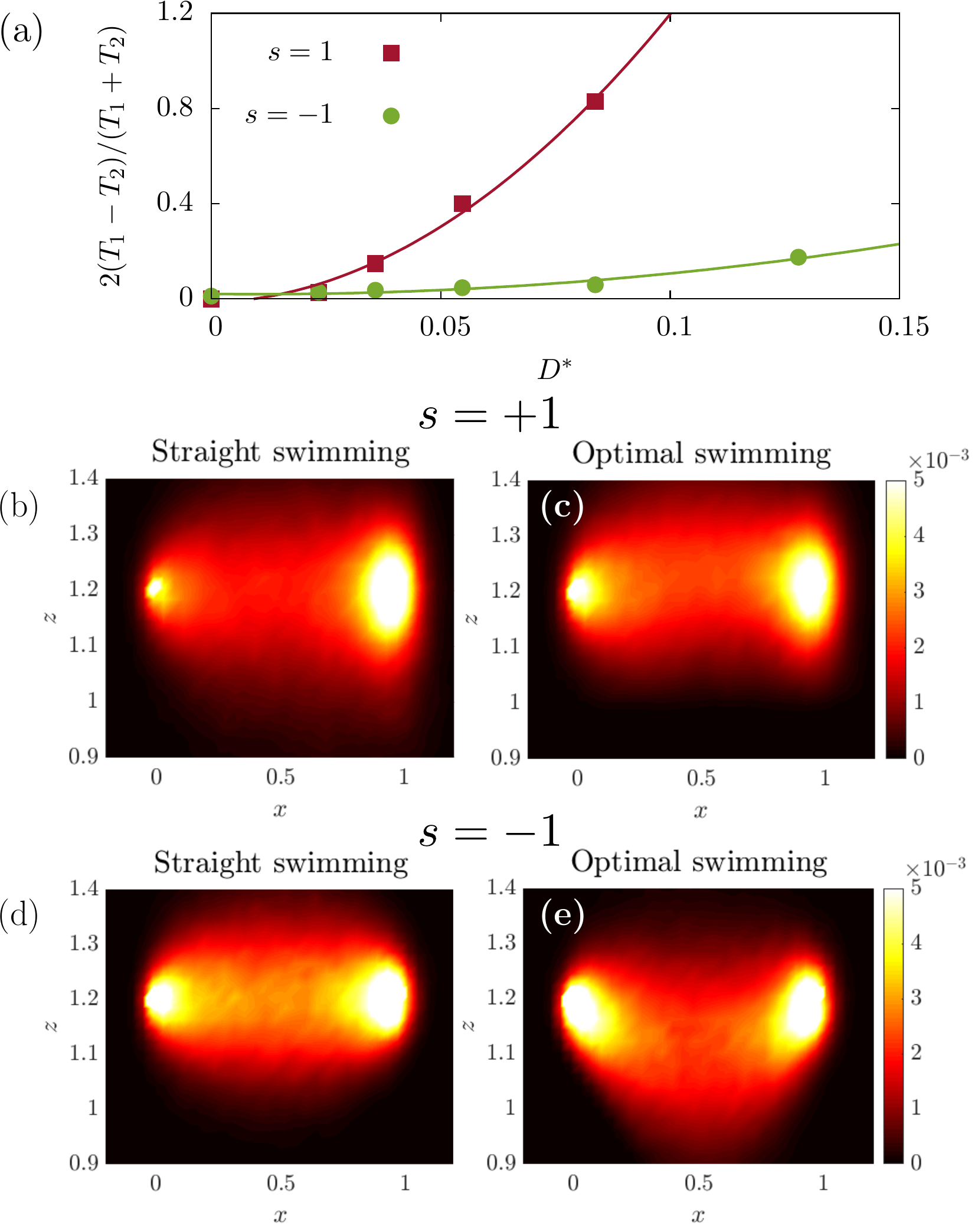}
\caption{Competition of navigation strategies in a fluctuating environment: 
(a) Relative travel-time difference between the straight swimming strategy (travel time $T_1$) and the optimal swimming strategy ($T_2$) for source dipole swimmers as a function of $D^\ast$.
{Panels} (b)--(e): Probability distribution (averaged over 5000 trajectories) of the coarse-grained position of a microswimmer navigating from 
{$\vect{r}_A=(0,0,1.2)$ to $\vect{r}_B=(1,0,1.2)$ for $D^\ast=0.025$ and $s=+1$ (b,c) or $s=-1$ (d,e).}
The target counts as reached when the microswimmer enters a spherical domain centered at the target point $\vect{r}_B$ of dimensionless radius of {$r^\ast=0.025$}. 
{Fluctuations are considered here in 3D.}
}
\label{fig4} 
\end{figure}

However, in our simulations we find that the optimal swimming strategy notably outperforms the straight swimming strategy over the entire considered noise regime ({Fig.~\ref{fig4}a}), i.e.\ from $D^\ast=0$ up to $D^\ast \approx 0.15$.
Interestingly, the difference between the two strategies increases with the noise strength,
such that the choice of the swimming strategy gets more and more important for a microswimmer as fluctuations become important. 
This finding might be relevant e.g.\ for microswimmers when trying to reach a food source: they do much better when seeking the proximity of nearby walls first (or getting into some reasonable distance), rather than greedily heading straight towards the target. 
\\To understand these observations, let us first 
consider the case $s=-1$ where
optimal swimming tends to reduce the microswimmer-wall distance and guides the swimmer to locations where hydrodynamic interactions are comparatively important and speed up the swimmer {(Fig.~\ref{fig4}d,e)}. Thus, for $s=-1$ the swimmer can steadily approach the target for comparatively large $D^\ast$-values. 
In contrast, when following the straight swimming strategy, nothing stops fluctuations from transferring the swimmer to 
regions where it is very slow ({Fig.~\ref{fig4}e}). 
The swimmer is then dominated by noise at comparatively low $D^\ast$-values and 
might reach the target only after following a long and winding path. 

Let us now discuss the case $s=1$, where optimal swimming reduces traveling times over the whole range of explored $D^\ast$-values, although 
the above mechanism does not apply, because swimmers slow down when they are close to the wall. 
To see the strategic advantage of optimal swimming also here, note that when following the straight swimming strategy, fluctuations may accidentally displace the swimmer to locations 
close to the wall, where it is slow. 
In contrast, the optimal swimming strategy makes the swimmer stay away from the wall ({Fig.~\ref{fig4}b,c}) and prevents it from getting trapped in regions where it is slow and dominated by noise.

\begin{figure}
	\centering
	\includegraphics[width=0.49\textwidth]{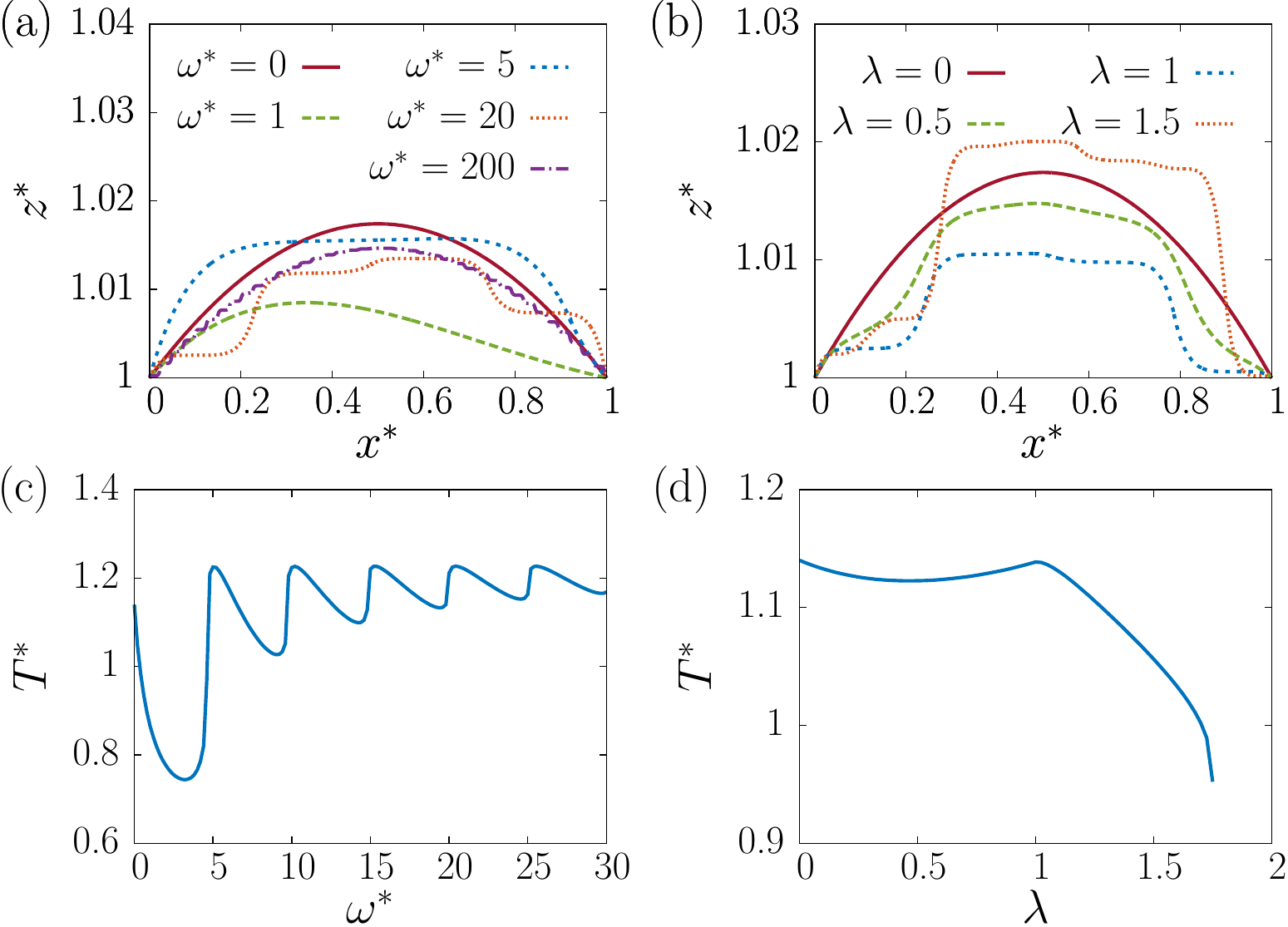}
	\caption{Optimal trajectories and travel times of time-dependent source-dipole microswimmers 
    for fixed amplitude $\lambda=1$ and different
    dimensionless frequencies 
	$\omega^\ast=\omega \tau=\omega v_0^{4/3}/|\sigma|^{1/3}$ shown in the key (a,c) and for fixed $\omega^\ast=6\pi$ and different $\lambda$ (b,d). 
	The shown travel times, $T^\ast=T/\tau=T v_0^{4/3}/|\sigma|^{1/3}$, for the optimal trajectory are always shorter than when choosing the shortest path to the target. 
	}
	\label{fig5}
\end{figure}

\vskip 0.5cm
\textbf{Time-dependent microswimmers:} 
We finally complement our discussion of the optimal microswimmer navigation problem by an exploration of time-dependent cases. 
This is inspired by microswimmers moving by body-shape deformations such as e.g. the algae {\it Chlamydomona reinhardtii}, 
which alternatively moves 
forward (stroke) and backwards (recovery stroke) and creates an oscillatory flow field~\cite{guasto2010}. 
We exemplary consider a time-dependent source dipole swimmer with 
$\dot x=[v_t-\sigma_t/(4z^3)]\cos\psi,\; \dot z=[v_t-\sigma_t/z^3]\sin\psi$, 
where $v_t:=v_0 g_1(t)$ and $\sigma_t:=\sigma g_2(t)$ are explicitly time-dependent functions. 
Using again length- and time-units 
$\ell = (\left| \sigma \right| /v_0)^{1/3}$, 
$\tau= \left| \sigma\right|^{1/3} /v_0^{4/3}$, 
this translates to
$\dot x^\ast=[g_1-s g_2/(4{z^\ast}^3)]\cos\psi,\; \dot z^\ast=[g_1-s g_2/{z^\ast}^3]\sin\psi$. 
While the case $g_1=g_2$ yields the same trajectories as the time-independent case (not shown), for
$g_1\neq g_2$ nontrivial trajectories occur (Fig.~\ref{fig5}a,b). Necessary conditions for these trajectories  
can be determined based on Pontryagin's maximum 
principle from optimal control theory~\cite{fuller1963,lee67,bryson1975,kirk2004} 
as detailed in the Methods section.
Choosing for illustrative purposes e.g. $g_1=1+\lambda \sin(\omega t)=1+\lambda \sin(\omega^\ast t^\ast)$, where $\omega^\ast=\omega \tau$, and $g_2=g_1^2/2$, 
leads to optimal trajectories (Fig.~\ref{fig5}a,b) which feature a characteristic step-plateau-like structure. Following such a trajectory the microswimmer mainly changes its distance from the wall in phases where 
it is slow, essentially to ``improve'' its distance from the wall for subsequent phases. 
When $\omega^\ast$ increases, the plateau length decreases and for $\omega^\ast \to \infty$ the optimal trajectory approaches a parabola (purple curve in panel (a)) which differs from the optimal trajectory for $\lambda=0$, because we have $\langle g_2(t)\rangle \neq \langle g_1(t)\rangle$ for the time-averages of $g_1,g_2$.

The resulting travel time, monitored as a function of frequency (Fig. \ref{fig5}c),
features a sequence of extrema 
occurring at frequencies where the swimmer reaches the target before completing a full driving cycle. For example, 
the global minimum corresponds to $\omega^\ast \approx \pi$ where the swimmer reaches the target at maximum speed
without experiencing a phase where the swimmer is slower than its average speed. 
The travel time also depends non-monotonously on $\lambda$; it features a 
local minimum around $\lambda=0.5$, where the time-average $\langle g_2(t)\rangle$ is smallest, 
and a local maximum at $\lambda=1$.
To understand the decrease of the travel time for $\lambda>1$, note that for $\lambda>1$  
the velocity temporarily changes sign. Since the swimmer can freely steer, it immediately turns and swims forward again with an effective speed of $v_0|1+\lambda \sin(\omega t)|$. This leads to an average swimmer speed which increases with 
$\lambda$, yielding the observed decrease of $T$.
These exact results exemplify the complexity of finding the optimal strategy in time-dependent cases and might serve as useful reference calculations to challenge corresponding machine-learning based approaches. 

\vskip 0.5cm 
\textbf{Parameter regimes:}
Let us now briefly discuss the generic relevance of our results for typical microswimmers. 
The force dipole coefficient of pushers and pullers is expected to scale as $\alpha \sim a^2 v_0$ \cite{desai2020biofilms,note1} where $a$ is the body size of the microswimmer. Thus, for $v_0 \sim 10~\mu$m/s we have $\ell \approx a$.
{Following Figs.~\ref{fig2}b, \ref{fig3}} this means that even at a wall distance of {2 -- 3 body length,} which commonly {occurs} in the life of many microswimmers, the deviation of the optimal path from the shortest one is significant. 
These estimates can be specified for E.\ coli bacteria where the 
force dipole coefficient has been measured in various experiments and amounts to 
$\alpha = 8-75~\mu$m$^3$/s \cite{desai2020biofilms} yielding $\ell=\sqrt{ |\alpha| /v_0}\sim 0.5-2~\mu$m for $v_0 \sim 20~\mu$m/s, which again is comparable to the length scale of the swimmer and means that the influence of hydrodynamic wall interactions on the required navigation strategy to reach a target fastest is highly significant. A similar discussion also applies to source dipole microswimmers, 
where the source dipole coefficient is expected to scale as $\sigma \sim a^3 v_0$. 
\\Regarding the generic relevance of our findings in the presence of noise, let us now estimate the typical value of $D^\ast=D/\left( |\sigma|^{1/3}v_0^{2/3} \right)$ for source dipole swimmers to compare with Fig.~\ref{fig4}a. Using $\sigma \sim a^3 v_0$ as well as the Stokes-Einstein relation $D=k_\mathrm{B} T/(6\pi \eta a)$, where $\eta \approx10^{-3}~$Pa~s is the viscosity of water and $k_\mathrm{B} T$ is the thermal energy, at room temperature, we obtain $D^\ast \sim 10^{-2}-10^{-1}$, 
depending on the size of the considered microswimmer. 
This roughly coincides with the parameter regime shown in Fig.~\ref{fig4}a, 
which means that typical microswimmers can save a significant fraction of their traveling time to reach a food source or another target lying one or a few body lengths away from a wall, when strategically regulating their distance to the wall rather than greedily heading straight towards the target.

\section*{Conclusions}
The message of this work is that, to reach their target fastest, microswimmers require navigation strategies which qualitatively differ from those used to optimize the motion of dry active particles or 
motile macroagents like airplanes. 
This finding hinges on hydrodynamic interactions between microswimmers and remote boundaries, which oblige the swimmers to take significant 
detours to reach their target fastest, even in the absence of external fields. 
Such strategic detours are particularly useful in 
the presence of (strong) fluctuations: they effectively protect microswimmers against fluctuations and allow them to 
reach a food source or another target up to twice faster than when 
greedily heading straight towards it. 
This suggests that strategically controlling their distance to remote walls might benefit the survival of motile microorganisms -- which serves as an alternative to the 
common viewpoint, that the microswimmer-wall distance is a direct (i.e.\ non-actively-regulated) consequence of hydrodynamic interactions.
\\Our results might be relevant for future studies on microswimmers in various complex environments involving hard walls or obstacle landscapes \cite{volpe2011microswimmers,brown2016swimming,dietrich2018active}, penetrable boundaries \cite{daddi2019membrane,daddi19njp}
or external (viscosity) gradients \cite{liebchen2018viscotaxis,laumann2019,datt2019active}.
For such scenarios our results (or generalizations based on the same framework) can be used as 
reference calculations e.g.\ to test machine learning based approaches to optimal microswimmer navigation~\cite{Yang2018,Yang2019} and perhaps also to help programming navigation systems for future microswimmer generations. They should also serve as a useful ingredient for future works on microswimmer navigation problems in environments which are not globally known but subsequently discovered 
by the microswimmers. 
Finally, for future work, it would also be interesting to explicitly solve the Hamilton-Jacobi-Bellman equation for the present problem with noise to compare the discussed navigation strategies 
which are optimal in the absence of noise and highly useful in the presence of noise, 
with the optimal navigation strategy following from this equation.

\fontsize{9}{11}\selectfont

\fontsize{9}{11}\selectfont

\section*{Methods}
Here we discuss details regarding the two approaches used to solve the optimal microswimmer navigation problem based on a Lagrangian approach and on Pontryagin's maximum principle respectively. 
Both approaches lead to identical results but have been found to be advantageous in different situations: the Lagrangian approach leads to a boundary value problem which is 
more immediate to implement, numerically simpler and more robust than the corresponding higher-dimensional problem resulting from 
Pontryagin's principle. The latter in turn allows for solving more general problems applying e.g.\ also to explicitly time-dependent microswimmers. 
\subsection*{Lagrangian approach for source dipole microswimmers:}
To find the optimal path we write the path connecting the starting point ($x_A^\ast,z_A^\ast$) and the terminal point ($x_B^\ast,z_B^\ast$) as a function {$z^*(x^*)$} and write the traveling time 
as
$T^\ast = \int_{x_A^\ast}^{x_B^\ast} |\dot x^\ast|^{-1} \,{\rm d} x^\ast = \int_{x_A^\ast}^{x_B^\ast} \mathcal{L}^\ast \left( x^\ast, z^\ast(x^\ast),{z^\ast}^\prime(x^\ast) \right) \, \Intd x^\ast$.
Following the Lagrangian optimization approach, a necessary condition for minimizing $T^\ast$
follows from the Euler-Lagrange equation
\begin{equation}
	\frac{\Intd}{\Intd x^\ast} \frac{\partial \mathcal{L}^\ast}{\partial {z^\ast}^\prime} - \frac{\partial \mathcal{L}^\ast}{\partial z^\ast} = 0 \, .
	\label{EL-Eq}
\end{equation}
where the Lagrangian, $\mathcal{L}^\ast$, depends on the microswimmer under consideration. 
{First, considering source-dipolar hydrodynamic interactions with a planar interface, an explicit expression for the Lagrangian can readily be obtained.
It follows from Eqs.~\eqref{sdeq} that $\cos\psi = x^* / \left( 1-s/ \left(4 {z^*}^3\right) \right)$ and $\sin\psi = z^* / \left( 1 - s/{z^*}^3 \right)$.
By enforcing the relation $\cos^2\psi + \sin^2\psi = 1$ and using the fact that $\dot{z}^* = {z^*}' \dot{x}^*$ with ${z^*}' = \partial z^* / \partial x^*$, the Lagrangian can explicitly be obtained as}
\begin{equation}
\mathcal{L}_\mathrm{SD}^\ast := \left| \dot{x}^\ast \right|^{-1} 
= \left( \left( 1-\frac{s}{4{z^\ast}^3} \right)^{-2} + {{z^\ast}^\prime}^2 \left(  1 - \frac{s}{{z^\ast}^3} \right)^{-2} \right)^{\frac{1}{2}} \, .
\end{equation}
{Inserting this Langrangian into the Euler-Lagrange equation~\eqref{EL-Eq}} shows that the optimal swimming trajectory is governed by the following second-order differential equation
\begin{equation}
	A(z^\ast) {z^\ast}''(x^\ast) + B(z^\ast) {z^\ast}'(x^\ast)^2 + C(z^\ast) = 0 \, , \label{SD-EL}
\end{equation}
where we have defined the coefficients
\begin{subequations}\label{SD-EL-Coeffs}
	\begin{align}
		A(z^\ast) &=  z^\ast(x^\ast) \left( z^\ast(x^\ast)^3-s \right) \left( 4 z^\ast(x^\ast)^3-s\right)^3 \, , \\
		B(z^\ast) &= -3s\left( 2 z^\ast(x^\ast)^3+s \right) \left( 4 z^\ast(x^\ast)^3-s \right)^2 \, , \\
		C(z^\ast) &= 48 s \left( z^\ast(x^\ast)^3 - s \right)^3 \, .
	\end{align}
\end{subequations}

Equations \eqref{SD-EL} and \eqref{SD-EL-Coeffs} subject to Dirichlet boundary condition of imposed vertical distance at the start and end points can readily be solved numerically using a computer algebra software by means of a standard shooting method.  

\subsection{Lagrangian approach for force dipole microswimmers:}
Next, for force-dipolar hydrodynamic interactions, we
first solve Eq.~\eqref{fdeqb} for the orientation angle~$\psi$, which yields four distinct solutions.
They are given by
\begin{subequations}\label{4psiSolutions}
	\begin{align}
		\psi_{1,2} &= \arctan \left( A_+, \pm B_+ \right) \, , \\
		\psi_{3,4} &= \arctan \left( A_-, \pm B_- \right) \, ,  
	\end{align}
\end{subequations}
where we have defined the arguments
\begin{subequations}
	\begin{align}
		A_\pm &= \frac{s}{9} \left( -4{z^\ast}^2 \pm \phi_1 \right) \, , \\
		B_\pm &= \frac{s}{9} \left( \phi_2 \pm 8 {z^\ast}^2 \phi_1 \right)^{1/2} \, , 
	\end{align}
\end{subequations}
with
\begin{subequations}
	\begin{align}
		\phi_1 &= \left( 72 s {z^\ast}^2 \dot{z}^\ast + 27 + 16{z^\ast}^4 \right)^{1/2} \, , \\
		 \phi_2 &= 2 \left( -36 s {z^\ast}^2 \dot{z}^\ast + 27 - 16{z^\ast}^4 \right) \, .
	\end{align}
\end{subequations}

Note that, for $a,b \in \mathbb{R}$, the function $\arctan(b,a)$ returns the principal value of the argument of the complex number $c = a + ib$, i.e.~\cite{abramowitz72},
\begin{equation}
	\arctan(b,a) = -i \ln \left( \frac{c}{|c|} \right) \in (-\pi, \pi] \, , 
\end{equation}
where $|c| = \left(a^2+b^2\right)^{1/2}$.
{We note that only real values of the steering angle should be considered.}
Now inserting Eqs.~\eqref{4psiSolutions} into Eq.~\eqref{fdeqa}, setting $\dot{z}^\ast = {z^\ast}^\prime \dot{x}^\ast$, and solving the resulting equations for $\dot{x}^\ast$, the Lagrangian is obtained as
\begin{equation}
	\mathcal{L}_\mathrm{FD} := \left| \dot{x}^\ast \right|^{-1}
	= \left| \frac{72 {z^\ast}^2 {z^\ast}^\prime}{r_\pm^2 - 16{z^\ast}^4 - 27 } \right| \, , 
	\label{Lagr-Dipole}
\end{equation}
where~$r_\pm$ are the roots of the quartic polynomial
\begin{equation}
	P_\pm(Z) = a_0 \pm a_1 Z + a_2 Z^2 +\pm a_3 Z^3 + a_4 Z^4 \, , \label{Polynom}
\end{equation}
the coefficients of which are explicitly given by 
	\begin{align}
		a_0 &= 9 \left( 16 {z^\ast}^4+27 \right)^2 + 256 {{z^\ast}^\prime}^2 {z^\ast}^4 \left(16{z^\ast}^4 -  81 \right) \, , \notag \\
		a_1 &= -64 {{z^\ast}^\prime}^2 {z^\ast}^2 \left( 16{z^\ast}^4 + 81 \right) \, , \notag \\
		a_2 &= -6 \left( 3+2 {{z^\ast}^\prime}^2 \right) \left( 16{z^\ast}^4 + 27 \right) \, , \notag \\
		a_3 &= 32 {{z^\ast}^\prime}^2 {z^\ast}^2 \, , \notag \\
		a_4 &= 9 + 4 {{z^\ast}^\prime}^2 \, . \notag
	\end{align}
The nature of the roots of the quartic polynomial is primarily determined by the sign of the discriminant~\cite{akritas89}.
Assuming that $z^\ast$ is a weakly-varying function about the value~$h>0$, such that $z^\ast(x^\ast) = h + \epsilon f(x^\ast)$, where $|\epsilon| \ll h$, the discriminants~$\Delta_\pm$ of the polynomial function given by Eq.~\eqref{Polynom} can be expanded to leading order about $\epsilon = 0$ as
\begin{equation}
	\Delta_\pm \sim K \left( 27 + 16h^4 \right) \left( 3s + 4h^2 \right)
	\left( 3s - 4h^2 \right) {{z^\ast}^\prime}^4 \, , \notag
\end{equation}
where $K$ is a positive real number.
In the far-field limit, we expect that $h^2 \gg 3/4$, and thus $\Delta_\pm < 0$.
Accordingly, the polynomial functions has two distinct real roots and two complex conjugate non-real roots~\cite{rees22}.

If we denote by $r_1$ and $r_2$ the real roots of $P_+$ then it can readily be noticed that $-r_1$ and $-r_2$ are the real roots of $P_-$ since $P_-(-Z) = P_+(Z)$.
Consequently, the system admits two possible Lagrangians, as can be inferred from Eq.~\eqref{Lagr-Dipole}.

The roots $r_1$ and~$r_2$ can be obtained via computer algebra systems.
They are not listed here due to their complexity and lengthiness.

Physically, the Lagrangian yielding the shortest traveling time is the one that needs to be considered~\cite{liebchen2019}.

\subsection{Lagrangian approach for force-quadrupolar microswimmers:}
Finally, we investigate the optimal swimming due to force-quadrupolar hydrodynamic interactions with the interface.
In this case, the translational swimming velocities read~\cite{spagnolie12, daddi18}
\begin{subequations} \label{Force-Quad-Gleischungen}
	\begin{align}
		\dot{x} &= v_0 \cos\psi + \frac{\nu \cos\psi}{32z^3} 
		\left[ 27\cos \left(2\psi\right) - 13 \right] \, , 
		\label{VX-Force-Quad} \\
		\dot{z} &= v_0 \sin\psi + \frac{\nu \sin\psi}{8z^3}
		\left[ 9 \cos \left(2\psi\right) + 5 \right] \, , 
		\label{VZ-Force-Quad}
	\end{align}
\end{subequations}
where~$\nu$ is the force quadrupolar coefficient.
In units of $\ell=(\left| \nu \right| /v_0)^{1/3}$ 
and $\tau= \left| \nu\right|^{1/3} /v_0^{4/3}$, Eqs.~\eqref{Force-Quad-Gleischungen} can be expressed in a dimensionless form as
\begin{subequations} \label{Force-Quad-Gleischungen-Scaled}
	\begin{align}
		\dot{x}^\ast &= \cos\psi + \frac{s \cos\psi}{32 {z^\ast}^3} 
		\left[ 27\cos \left(2\psi\right) - 13 \right] \, , 
		\label{VX-Force-Quad-Scaled} \\
		\dot{z}^\ast &= \sin\psi + \frac{s\sin\psi}{8 {z^\ast}^3}
		\left[ 9 \cos \left(2\psi\right) + 5 \right] \, , 
		\label{VZ-Force-Quad-Scaled}
	\end{align}
\end{subequations}
where we have used the abbreviation $s=\operatorname{sgn}(\nu)$.

Solving Eq.~\eqref{VX-Force-Quad-Scaled} for~$\psi$ yields three possible distinct values 
\begin{subequations}\label{psiSols}
	\begin{align}
		\psi_1 &= \arccos	\left(  \phi_{+} \right) , \label{psi1} \\
		\psi_2 &= \pi - \arccos \left(  \frac{1}{2} \, \phi_{+} - i \, \frac{\sqrt{3}}{2} \, \phi_- \right)  , \\
		\psi_3 &= \pi - \arccos \left( \frac{1}{2} \, \phi_{+} + i \, \frac{\sqrt{3}}{2} \, \phi_- \right)  ,
	\end{align}
\end{subequations}
where we have defined for convenience the abbreviations $Z = {z^\ast}^3/s$ and $E = \left( 27\dot{x}^\ast Z + f^{\frac{1}{2}} \right)^{\frac{1}{3}}$.
Moreover,
\begin{equation}
	f = 64Z^3 + 3 \left( 243\dot{x}^2 - 80 \right)Z^2 + 300Z - 125 \, , 
\end{equation}
and 
\begin{equation}
	\phi_\pm = \frac{2}{9} \left( E \pm \frac{5-4Z}{E} \right) \, .
\end{equation}

Substituting the expression of~$\psi_1$ given by Eq.~\eqref{psi1} into Eq.~\eqref{VZ-Force-Quad-Scaled}, using the fact that $\dot{z}^\ast = {z^\ast}^\prime \dot{x}^\ast$, and solving the resulting equation for $\dot{x}^\ast$ yields the expression of the Lagrangian
\begin{equation}
	\mathcal{L}_\mathrm{FQ} := \left| \dot{x}^\ast \right|^{-1}
	= \frac{54 {z^\ast}^3}{\left| \rho^3 - s \rho^{-3} \left(4 {z^\ast}^3-5s\right)^3 \right|} \, , 
\end{equation}
where~$\rho$ denotes the roots of explicitly-known polynomial of degree~12.
In the physical range of parameters, this polynomial admits either one real radical having the order of multiplicity four or three distinct radicals also having the order of multiplicity four.

It turned out that the same set of radicals are obtained when making substitution with $\psi_2$ of $\psi_3$.
{Again, only real values of the steering angle should physically be considered.}

In order to proceed further, we evaluate numerically the Lagrangians and accurately fit the results using a standard nonlinear bivariate hypothesis of the form
\begin{equation}
	\mathcal{L} (z^\ast, {z^\ast}^\prime) = \sum_{m=0}^{N} \sum_{n=0}^N a_{ij} {z^\ast}^m {{z^\ast}^\prime}^n \, , 
\end{equation}
where~$a_{ij}$ are fitting parameters.
Here, we have taken {$N=5$} but checked that taking larger values does not alter our results.

\vskip 0.5cm
\textbf{Hydrodynamic interactions near spherical boundaries:}

\begin{figure}
	\centering
	\includegraphics[scale=1.2]{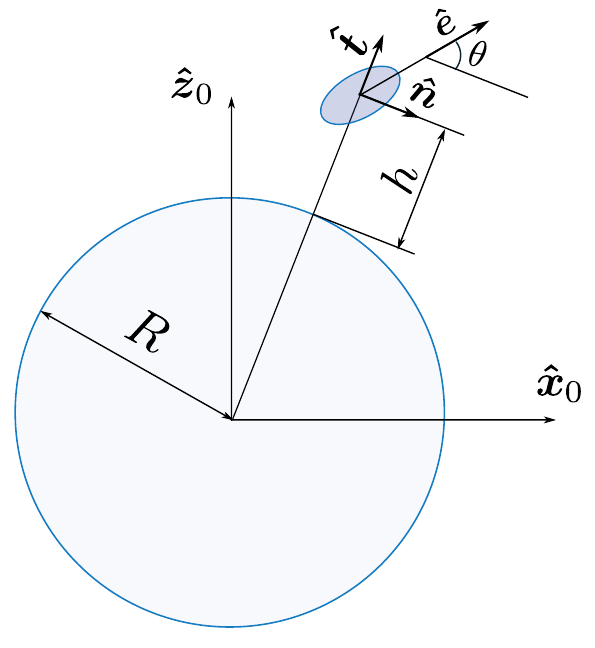}
	\caption{Graphical illustration of a microswimmer moving near a spherical boundary.}
	\label{Illus-Sph}
\end{figure}

The translational swimming velocities resulting from source dipolar hydrodynamic interactions with a solid sphere of radius~$R$ positioned at the origin of coordinates (see Fig.~\ref{Illus-Sph}) can be decomposed into two terms
\begin{equation}
	\vect{V}  = \vect{\hat{e}} + \vect{v}^\mathrm{HI} \, , 
	\label{SD-Main}
\end{equation} 
with $\vect{\hat{e}} = \cos\theta \, \vect{\hat{n}} + \sin\theta \, \vect{\hat{t}} $ denoting the instantaneous orientation angle of the swimmer, and 
\begin{equation}
	\vect{v}^\mathrm{HI} =  P \sin\theta \, \vect{\hat{t}} + Q \cos\theta \,  \vect{\hat{n}} \, , 
\end{equation}
quantifies the effect of hydrodynamic interactions with the spherical boundary.
This contribution can readily be determined from the Green's function near a rigid sphere~\cite{spagnolie15}.
Here, we have defined for convenience
\begin{subequations}
	\begin{align}
		P &= -\frac{\sigma R \left( 3h^2 + 6hR + 8R^2\right)}{h^3 \left(h+2R\right)^3} \, , \\
				Q &= \frac{\sigma R \left( 3h^2+6hR+4R^2 \right) \left( h^2+2hR-2R^2 \right)}{4h^3 \left( h+2R \right)^3 \left(h+R\right)^2} \, ,
	\end{align}
\end{subequations}
where, again, we have scaled lengths by a characteristic length scale of the swimmer~$L$, and velocities by the bulk swimming speed~$v_0$.

Without loss of generality, we consider motion in the plane~$y=0$.

To obtain the translational swimming velocities near two obstacles, as illustrated in Fig.~\ref{fig1}~a of the main text, we use the commonly-employed superposition approximation \cite{daddi2020axisymmetric, dufresne2001brownian}.
The latter conjectures that the solution for the Green's function near two widely-separated obstacles can conveniently be approximated by superimposing the contributions due to each obstacle independently.
Accordingly,
\begin{equation}
	\vect{V} = \vect{\hat{e}}_1 + \vect{v}^\mathrm{HI}_1 + \vect{v}^\mathrm{HI}_2 \, , \label{V2sph}
\end{equation}
where
\begin{equation}
	\vect{v}^\mathrm{HI}_i =  P_i \sin\theta_i \, \vect{\hat{t}}_i 
	+ Q_i \cos\theta_i \,  \vect{\hat{n}}_i \, , \quad i \in \{1,2\} \, .
\end{equation}
In addition,
\begin{equation}
	\cos\theta_1 \, \vect{\hat{n}}_1 + \sin\theta_1 \, \vect{\hat{t}}_1 
	=: \vect{\hat{e}}_1 \equiv \vect{\hat{e}}_2 :=
	\cos\theta_2 \, \vect{\hat{n}}_2 + \sin\theta_2 \, \vect{\hat{t}}_2 \, . \label{e2sph}
\end{equation}
We then project Eq.~\eqref{V2sph} along the unit vectors~$\vect{\hat{t}}_1$ and~$\vect{\hat{n}}_1$ to obtain two scalar equations.
Next, we project Eq.~\eqref{e2sph} along the unit vectors~$\vect{\hat{t}}_2$ and~$\vect{\hat{n}}_2$ to obtain two additional equations.
Subsequently, the unknown quantities $\cos\theta_1$, $\sin\theta_1$, $\cos\theta_2$, and $\sin\theta_2$ can be expressed in terms of~$x$, $z$, and~$z'$ by solving the resulting linear system of four equations, upon using the fact that $\dot{z} = z^\prime \dot{x}$.
By solving the identity $\cos^2\theta_1 + \sin^2\theta_1 = 1$ for~$\dot{x}$, the Lagrangian can then be obtained explicitly as $\mathcal{L}_\mathrm{SD} = \left| \dot{x} \right|^{-1}$.
The expression of the resulting Lagrangian is rather lengthy and complicated and thus omitted here.
Finally, the Euler-Lagrange equation can be solved numerically in Matlab using the {\ttfamily ode45} routine.
\subsection{Optimal control theory for microswimmer navigation:}
We now use Pontryagin's principle from optimal control theory~\cite{fuller1963,lee67,bryson1975,kirk2004}
as an alterntive approach to solve the optimal navigation problem for microswimmers. 
While the 
Lagrangian approach is numerically rather convenient for the present set of problems, 
Pontryagin's approach provides a more general framework to determine optimal 
microswimmer trajectories. In particular, it 
allows us to discuss optimal trajectories in higher spatial dimensions or to calculate the optimal path for microswimmers with a time-dependent propulsion speed as we discuss 
now. 

Let us consider the cost 
$J[\vect r(t),\vect p(t),\psi(t),t] = T$ 
which we want to minimize 
subject to the equations of motion for microswimmers interacting with remote walls {(Eqs.~\eqref{sdeq})} and 
the boundary conditions $\vect{r}(t=0)=\vect{r}_A$ and $\vect{r}(t=T)=\vect{r}_B$ where $\vect{r}:=(x,z)$. Here $\psi$ is the control variable, $\vect p$ is the costate and $T$ is the 
(unknown) traveling time corresponding to the fastest route. As the cost can be determined from the endpoint (or the ``endtime'') alone it can be considered as a pure endpoint cost, whereas the running cost is zero. 
Allowing for explicitly time-dependent microswimmers with a time-dependent source dipole strength $\sigma_t$ {and/or} 
a time-dependent self-propulsion speed $v_t$, this is a non-autonomous, fixed endpoint, free endtime optimal control problem. 
To solve it we first construct the optimal control Hamiltonian 
as $H({\vect{r}},{\vect{p}},\psi,t) = {\vect{F}}\cdot {\vect{p}}$ 
where $\vect F$ is given by $\dot {\vect r} =\left([v_t - \sigma_t/\left(4 z^3 \right)]\cos \psi ,[v_t - \sigma_t/z^3]\sin \psi \right)=: \vect F$. We now minimize $H$ with respect to the (unconstrained) control variable $\psi$ by solving $\partial_\psi H=0$ 
yielding $\tan \psi^\ast = u_z/u_x$ where $\psi^\ast$ is the optimal control and 
where we have defined $u_x:=p_x (v_t-\sigma_t/(4z^3))$ and $u_z=p_z (v_t-\sigma_t/z^3)$. Using $\mathcal{H}(\vect r, \vect p,t)\equiv H({\vect{r}},{\vect{p}},\psi^\ast,t)$, after a few lines of straightforward 
algebra, we obtain the minimized (lower) Hamiltonian as $\mathcal{H}(\vect r, \vect p,t)=\pm \sqrt{u_x^2+u_z^2}$. In the following we use only the plus--branch, as the minus--branch does not seem to result in sensible trajectories. 
The optimal trajectory now follows from the Hamilton equations of motion $\dot {\vect{r}}=\partial_{\vect{p}} \mathcal{H}$ and 
$\dot {\vect{p}}=-\partial_{\vect{r}} \mathcal{H}$ which have to be solved as a 
boundary value problem, 
such that ${\vect{r}}(t=0)={\vect{r}}_A$ and ${\vect{r}}(t=T)={\vect{r}}_B$ where $T$ is the unknown traveling time. 
Thus, up to now, we have determined four first order differential equations which are complemented by 
four boundary conditions. Since this problem depends on the unknown $T$, we need 
one additional boundary condition. This is given by 
the Hamiltonian endpoint condition (or transversality condition) which determines the value of $\mathcal{H}(\vect r(T),\vect p(T),T)$. This condition can be determined from the endpoint Lagrangian $\bar L= T + \vect \nu \cdot \vect e$ where $T$ is again the endpoint cost; $\vect \nu$ is the (constant) endpoint costate and $\vect e=({\vect{r}}(T)-{\vect{r}}_B)$ (such that $\vect e=\vect 0$ represents the boundary conditions in standardized form). The Hamiltonian endpoint condition simply yields $\mathcal{H}(\vect r(T),\vect p(T),T)=-\partial_T \bar L=-1$ for the considerd minimal time problem. 
(The remaining two transversality conditions 
for the costate, $\vect p(T) = -\partial_{\vect r(T)} \bar L$, just relate the two unknowns $\vect p(T)$ to two other unknowns $\vect \nu$ and hence do not provide any additional information.)
\\To numerically solve the Hamilton equations together with ${\vect{r}}(0)={\vect{r}}_A$, ${\vect{r}}(T)={\vect{r}}_B$ and $\mathcal{H}(\vect r(T),\vect p(T),T)=-1$, it is convenient 
to rescale time via $t=T t'$ such that, in primed units, the terminal time is $T=1$ and the endpoint boundary conditions 
are explicitly known as ${\vect{r}}(t'=1)={\vect{r}}_B$; i.e.\ they no longer depend on the unknown variable $T$, which instead shows up in the equations of motion. 
In addition, to ease the usage of standard shooting methods to solve this boundary value problem, it is convenient to treat $T$ 
as a dynamical variable, i.e.\ we introduce $T=:a(t)$, leading to the additional equation of motion $\dot a=0$, which we solve together with the Hamilton equations and the five mentioned boundary conditions. 
We have compared the results from this approach for time-independent microswimmers in the presence of flat walls with those from the Lagrangian approach and find the same trajectories. For time-dependent microswimmers corresponding new results are presented in the main text. 
\\Let us finally mention that an alternative approach to treat the present non-autonomous optimal control problem would be to first transform it to an autonomous problem, by introducing an extra variable $\dot \tau=1$ with $\tau(t=0)=0$ such that $\tau(t)=t$. The corresponding approach leads to the same numerical results to those presented in Fig.~\ref{fig5}.

\section*{Data availability}
The data that support the findings of this study are available from the authors upon reasonable request.

\section*{Code availability}
The code that supports the findings of this study are available from the authors upon reasonable request.

\input{main.bbl}

\section*{Acknowledgments} 
We thank A. M. Menzel for useful discussions and gratefully acknowledge
support from the DFG (Deutsche Forschungsgemeinschaft) through projects
DA~2107/1-1 and LO~418/23-1.

\vspace{0.5cm}
\section*{Author contributions}
H.L. and B.L. have planned the project. A.D.M.I. and B.L. have performed the analytical calculations; A.D.M.I. has 
performed the numerical simulations and prepared the figures. All authors have discussed and interpreted the results. B.L. and A.D.M.I. have written the main text and the methods section, respectively, with input from H.L.

\section*{Competing interests}
The authors declare no competing interests.

\section*{Additional information}
Correspondence and requests for materials should be addressed to B.L.

\end{document}

%% file: main.bbl
%

%% file: main.bbl
\begin{thebibliography}{79}
\bibitem [{not()}]{note1}%
  \BibitemOpen
  \href@noop {} {}\bibinfo {note} {Consider microswimmers which use beating
  elements (limbs) with an overall extension of $\ell$ to create a flow field
  which advects the body of size $a$ with a speed $v_0$. If the extension of
  the limbs is of the same order of magnitude as the body size, i.e. $\ell \sim
  a$, we obtain $\ell=\sqrt{|\alpha|/v_0}\sim a$ and hence $\alpha \sim a^2
  v_0$.}\BibitemShut {Stop}%
\end{thebibliography}
